\documentclass[pre,twocolumn, amsmath, amssymb, superscriptaddress]{revtex4}

\usepackage{braket}
\usepackage{subfigure}
\usepackage{adjustbox}
\usepackage{xcolor}
\usepackage{todonotes}
\usepackage{hyperref}

\begin{document}

\title{Genome heterogeneity drives the evolution of species}

\author{Mattia Miotto\footnote{The authors contributed equally to the present work.}}
\affiliation{Department of Physics, Sapienza University of Rome, Rome, Italy}
\affiliation{Center for Life Nanoscience, Istituto Italiano di Tecnologia,  Rome, Italy}

\author{Lorenzo Monacelli\footnotemark[1]}
\affiliation{Department of Physics, Sapienza University of Rome, Rome, Italy}

\newcommand{\eqname}{{Eq.}}
\newcommand{\secname}{Sec.}
\newcommand{\beq}{\begin{equation}}
\newcommand{\eeq}{\end{equation}}
\newcommand{\red}[1]{\textcolor{red}{#1}}
\newcommand{\blue}[1]{{\bf\textcolor{blue}{#1}}}

\begin{abstract}
Most of the DNA that composes a complex organism is non-coding and defined as \textit{junk}. Even the coding part is composed of genes that affect the phenotype differently. Therefore, a random mutation has an effect on the specimen fitness that strongly depends on the DNA region where it occurs. Such heterogeneous composition should be linked to the evolutionary process. However, the way is still unknown.

Here, we study a minimal model for the evolution of an ecosystem where two antagonist species struggle for survival on a lattice. Each specimen possesses a toy genome, encoding for its phenotype. The gene pool of populations changes in time due to the effect of random mutations on genes (entropic force) and of interactions with the environment and between individuals (natural selection).

We prove that the relevance of each gene in the manifestation of the phenotype is a key feature for evolution. In the presence of a uniform gene relevance, a mutational meltdown is observed. Natural selection acts quenching the ecosystem in a non-equilibriumstate that slowly drifts, decreasing the fitness and leading to the extinction of the species. Conversely, if a specimen is provided with a heterogeneous gene relevance, natural selection wins against entropic forces, and the species evolves increasing its fitness. We finally show that heterogeneity together with spatial correlations is responsible for spontaneous sympatric speciation.
\end{abstract}

\maketitle

Evolution by natural selection shaped the marvelous biodiversity we presently observe in nature.  
Starting with one (or few) living organisms, the tree of life progressively branched as living beings managed to survive to different environments through adaptations and speciations~\cite{howard1998endless}. 

Both processes arise by a complex interplay between intrinsic forces (e.g. mutations) and extrinsic ones, provided by the interactions between the different components of the ecosystem~\cite{COOK1906}. 
This results in changes of behavior, morphology, and physiology (or combinations thereof) in the organisms.

Since its first formulation, the theory of evolution regarded natural selection, i.e. the effect of environment and interaction inter and intraspecies in selecting the organisms with maximum fitness, as the pivotal mechanisms of evolution.
Several more years were required to clearly define the 'microscopic' role of genome mutations (alongside genetic drift, hitchhiking ext.) as the other main component of the evolutionary process~\cite{FREESE1965,mayr2004,Nei2007,Kimura1983}.

Quantifying the simultaneous effects of natural selection and genomic mutations is far from being an easy task~\cite{Eigen1971,Nei2007,Hershberg2015,Forcelloni2019}. Its understanding is however not only a fascinating theoretical challenge but carries important practical implications. In fact, a continuously increasing literature highlights the connection between the rules governing ecology to the one regarding cell populations~\cite{Zhang2012, Korolev2014,McGranahan2017}. Preeminent is the case of cancer cells, that are both subject to high entropic forces (fast mutation rates) and a strong natural selection, due to the selective pressure given by both the competition with the immune system and the effect of anti-cancer therapies. Bacteria drug resistance is under much scrutiny, too, since the selection done by antibiotics influences the evolution of resilient traits.
Experimentally, the evolution of genomes during speciation has been studied only recently with the availability of next-generation sequencing technologies~\cite{Cavalieri2000, Shapiro_2012,Campbell2018}.
In particular, recent findings highlighted how mutations on certain genes considerably enhance the speciation process~\cite{Campbell2018}.
Also the increase of the genome length, due to duplication errors (e.g. the presence of redundant genes/chromosomes) has been pointed as one mechanism for evolution/speciation~\cite{genomesize}.\\

Parallel to experiments,  several theoretical models were developed to study ecosystem dynamics\cite{Lotka,volterra} and evolution, both at  the molecular level~\cite{Eigen1971,Baji__2018} and at the population level, such as population-environment interaction~\cite{ExpExp, Kussell2005, 2DeMartino2017,Bena2007} and  specie-species interactions~\cite{Savill1998, HE2003, HE2005, Pkalski2006,P_rez_Escudero_2016, Abernethy2018,Barbier_2018,Portalier_2018}.

In particular, agent-based models proved to be very efficient to include spatial information/features \cite{dewdney, Antal2001, EcoLat, Mobilia2006, Dobramysl_2017}. Here, evolution is accounted either at phenotype level, i.e. the phenotype of a species is randomly modified generation by generation according to a particular law~\cite{Savill1998,Kussell_2005,ExpExp}, or at genotype level, i.e. an evolution law is assumed for a genome upon which the phenotype is computed~\cite{Moran_1976,Swetina_1982,Penna1995,Friedman_2013}.
 More refined models have been considered accounting for dominant and recessive alleles, sex recombination (cross-over) or spatially resolved ecosystem dynamics~\cite{Sousa2004,Friedman_2013,Dalmau2018}.

This work aims to study how heterogeneity in gene relevance affects the evolution of the species.
In particular, we study the evolution of a minimal ecosystem on a lattice, initially composed by one species of predators and one of preys.
Each specimen possesses a toy genome composed of $3N$ genes that encode for three essential macro-phenotypic features of the animal: i) its capability of moving/hunting, ii) its fertility and iii) the mean life-time of the specimen (mortality). 

The genome pool of the two prey-predator species can change in time due to random mutations at the level of the single genes (entropic force) and to predation/death events (selection force). 
Aiming at limiting the number of parameters and developing an evolutionary model as minimal as possible, we do not insert mating in the model, i.e. we have an asexual reproduction. 
Since mating is not taken into consideration, we will not look for speciation according to Mayr's Biology Speciation Concept, where different species are separated if mating does not produce fertile off-springs~\cite{mayr2004}. Instead, we speak of speciation in terms of differences in phenotype distributions~\cite{GarciaBernardo2016}, where well-separated peaks can be interpreted as different species while the width of each peak accounts for the intra-species differentiation. 

Two different kinds of genome are considered: i) a uniform genome (where all the genes have the same impact on the phenotype), and a heterogeneous genome (where each gene has a different weight on the overall phenotype). We show that, while in the first case entropy dominates, fostering the mutational meltdown~\cite{OHTA_1973,Lynch_1995,Grande_Perez_2005}, in the latter one natural selection allows the ecosystem to increase predator fitness. For each genome, we also look for the emergence of spontaneous speciation.


\begin{figure*}[t]
\includegraphics[width=12.4cm,height=14.4cm]{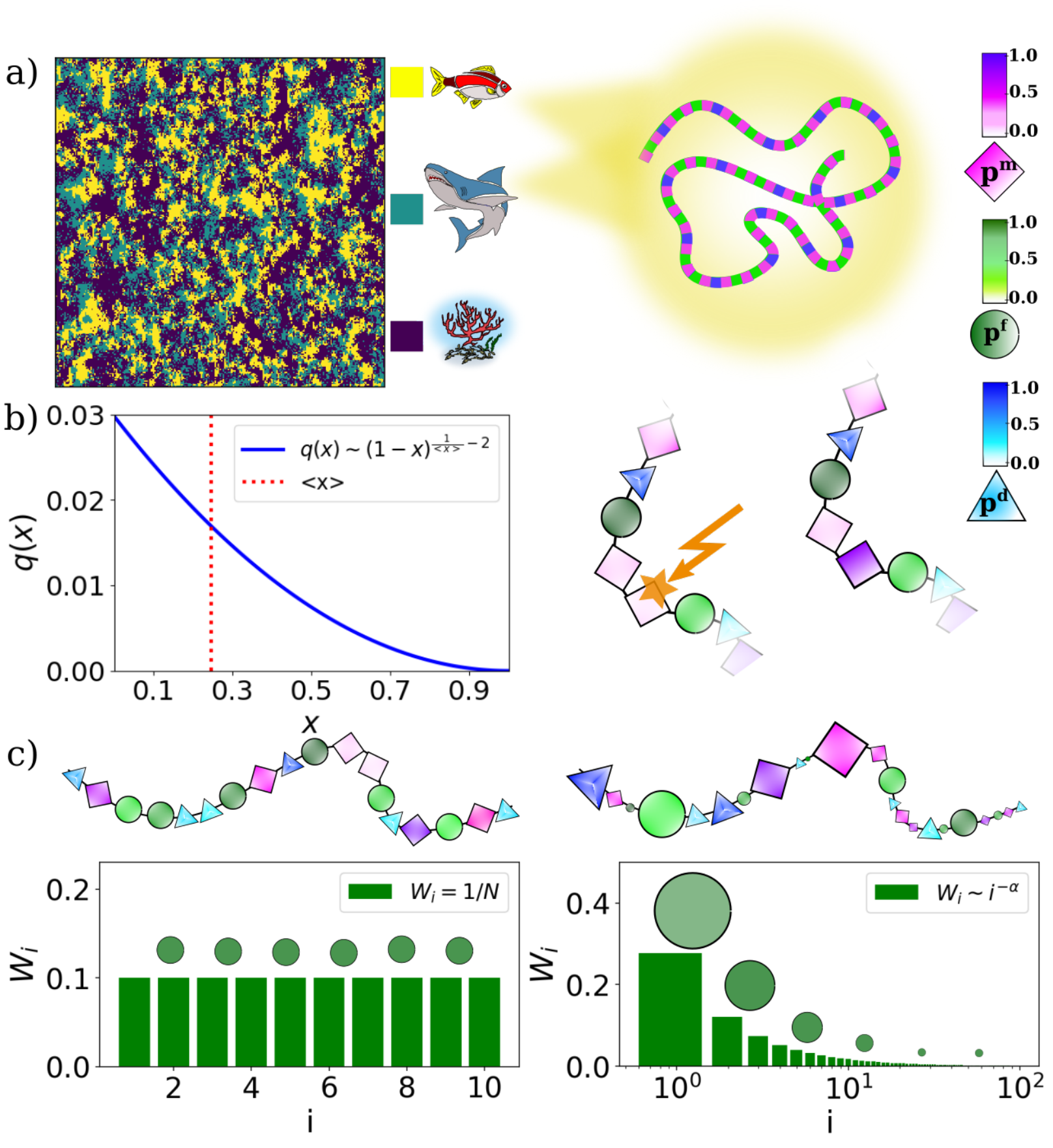}
\caption{\textbf{Features of the EvoLat model.} \textbf{a)} Snapshot of an EvoLat configuration in the steady-state regime. Fishes are colored in yellow, sharks in blue, while purple cells represent the environment. Each animal on the lattice possesses a toy genome, composed of $3N$ genes. Each gene is represented by a real number between 0 and 1 that concurs in manifesting three macro-phenotypes associated with the animals. Animal mobility is represented by the probability of moving, $p^m$, their fertility by the breeding probability ($p^f$) and their life-time by the death probability ($p^d$). \textbf{b)}Mutation events can happen with a rate, $\mu$ with equal probability on each gene. If one gene is subject to a mutation, a new value between 0 and 1 is extracted from the underline phenotype distribution $q(x)$, whose asymmetry accounts for the fact that mutations tend to be deleterious for the organisms. 
  \textbf{c)} The phenotype of the individual is obtained as the weighted average over all the $N$ independent genes encoding for it. In the present work, we assume two possible kinds of weights, uniform mean (all genes equally concur to the phenotype) or power-law weighted mean (some genes influence the phenotype more than others).
}
\label{fig:model}
\end{figure*}

\section{Model}
\label{sec:model}
We consider a variant of the EcoLat (Ecosystem on Lattice) model~\cite{EcoLat}, where each site of a $L\times L$ square lattice can be occupied exclusively either by the environment or a prey or a predator. 
To use the same notation adopted in~\cite{EcoLat}, we identify preys as fishes ($f$) and predators as sharks ($s$). 
At every discrete time step, fishes can move, breed or remain still with probability $p_f^m$, $p_f^f$ and $1 - p_f^m - p_f^f$. Sharks can move ($p_s^m$) or remain still ($1 - p_s^m$). Furthermore, predators eat preys whenever they step into a cell occupied by a prey. In this case, sharks can reproduce with probability $p_s^f$. If a shark does not eat a fish during its round, it can die for starvation with probability $p_s^d$. 
We assumed that fishes may only die murdered by a predator (i.e. $p_f^d = 0$).

The set of three probabilities, $\{p_{(s,f)}^m, p_{(s,f)}^f,p_{(s,f)}^d\}$, constitutes the macro-phenotype of each individual, dictating the rates with which the specimen carries out three essential tasks of life. 

Each specimen has a genome that codes the macro-phenotype of the individual.
The macro-phenotype is obtained as a weighted averages over all the $N$ genes ($g_{i}$) that code for a particular feature:
\beq
\label{eq:pheno}
p_y = \sum_{i=1}^N g_{yi} W_i\qquad y = {}_{(s,f)}^{(m,f,d)},
\eeq  
where $W_i$ expresses the weight of the $i$-th gene in the manifestation of the phenotype (see Figure~\ref{fig:model}\emph{a}).

\begin{figure*}
\includegraphics[width=14.4cm,height=10.4cm]{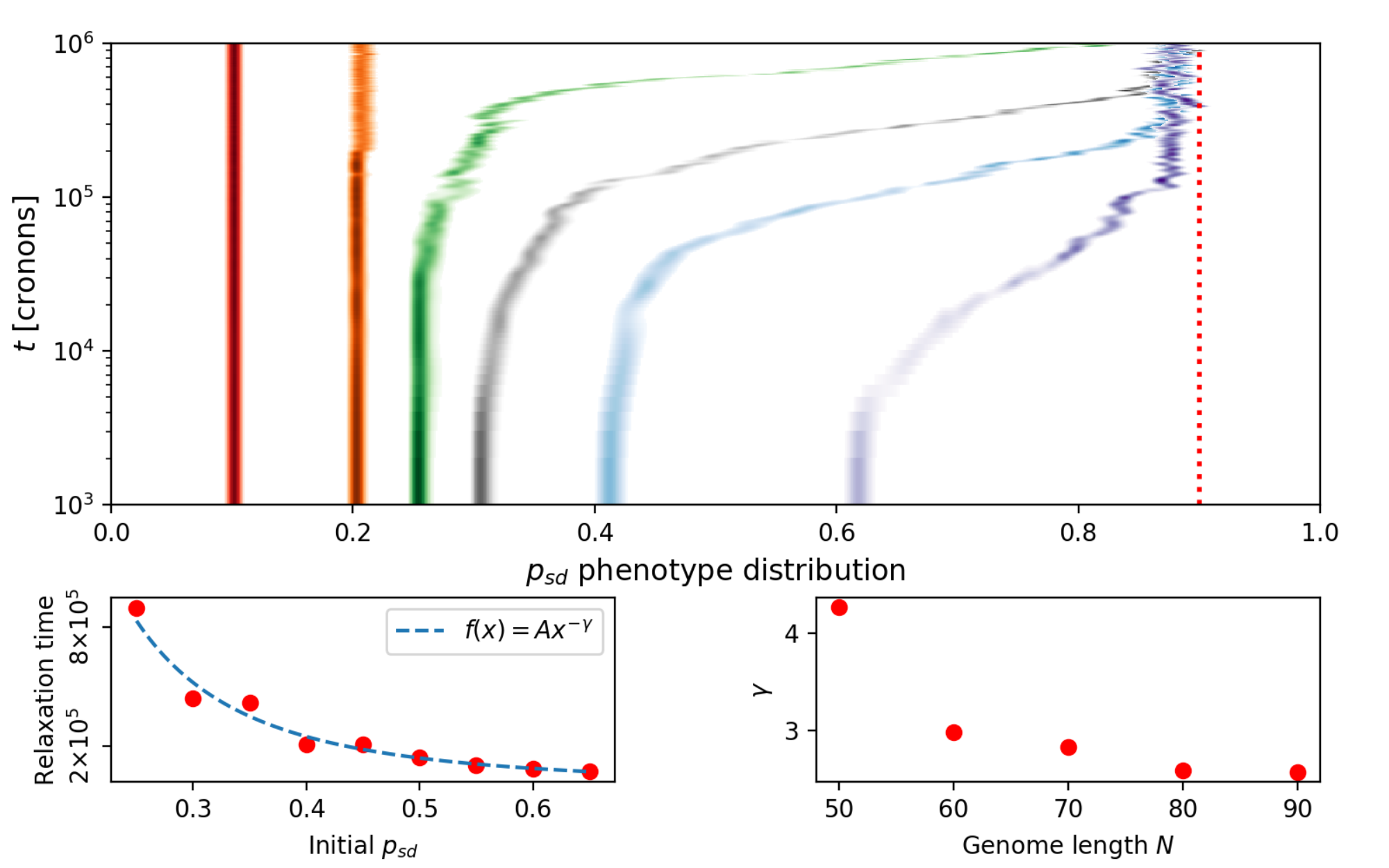}
\caption{\textbf{Time evolution of predators death-rate after quenching. The predator relaxation time diverges as a power-law, in a way that resemble the dynamic of glassy systems. Different colors correspond to different simulations performed starting from several values of $p_{sd}$.} \textbf{a)} Shark mortality distribution. The dashed vertical line is the stationary state for predator growing in a breeding farm (no natural selection).\textbf{b)} Relaxation time as a function of the initial phenotype. \textbf{c)} Exponent of the power law as a function of the genome length. It converges to a value of about 2.57. Simulation in panel $a$ and $b$ are performed with $N=90$.}
\label{fig:uniform:genome}
\end{figure*}
   
Genome is subject to point mutations, i.e. Poisson-distributed random events in time occurring with a constant rate for each gene every time a new individual is born (reproduction is asexual). 
If a mutation event occurs on a gene, its value is reset to a random number between 0 and 1. 
In order to tune the average value of the new mutated gene, we choose the new value $g_i = x$ according to the following probability distribution:

\beq 
\label{eq:q}
q(x) = \left( \frac{1}{\left<x\right>}-1\right) \left(  1- x\right)^{ \frac{1}{\left<x\right>}-2}
\eeq

with $\braket{x}$ is a tunable parameter that determines the average value of the gene after the mutation (see Figure~\ref{fig:model}\emph{b}). Such kind of asymmetric distributions is widely used for phenotype modeling, e.g. in the contest of bacterial growth rates, where slow rates dominate over fast ones~\cite{Martino_2016, DeMartino2017,ExpExp}. 
In this way, deleterious mutations can be enforced to be more common than favorable ones. This is quite reasonable, as a random mutation in the coding DNA produces a random change in the amino-acid chain of a protein, that is far more likely to produce an unfolded structure than a more functional one~\cite{Cocco_2018, thermo2018}.

Finally, in this work we will consider two weight distributions (see Figure~\ref{fig:model}\emph{c}): 
\begin{itemize}
    \item a uniform distribution, where all genes equally contribute to manifest the phenotype (discussed in Sec.~\ref{sec:1})
\begin{equation}
    \label{eq:uniform}
    W_i = \frac{1}{N},
\end{equation}

\item    a power-law distribution allowing for the presence of preeminent genes, as discussed in  Sec.~\ref{sec:2},
\begin{equation}
    \label{eq:power:law}
    W_i \propto i^{-\alpha}.
\end{equation}
\end{itemize}


To assess the outcome of evolution, we define the fitness measured as the average size of the population. Such definition, profoundly linked with the usual fitness measurements (growth rate or the number of nephews per individual), is enforced by the finite carrying capacity, dictated by the lattice.

\section{Results}\label{sec:res}

\subsection{Uniform genome}
\label{sec:1}

To begin with, we investigated the model behavior in presence of a uniform genome, where each phenotype is determined as the uniform average over all the genes associated to that phenotype (as given by \eqref{eq:pheno} with ~\eqref{eq:uniform}).

After preparing the ecosystem in a non-equilibrium steady state (NESS) with fixed initial phenotypes, we turned on mutations and followed the evolution of prey's and predator's traits on long periods.
To prevent predators to acquire infinite lifetimes, we strongly favored deleterious mutations in the shark death rate (entropic force). This is achieved by tuning the $\braket x$ parameter in \eqref{eq:q}. In fact, if the average phenotype is above (or below) $\braket x$, random mutations will try to restore it around $\braket x$.

\figurename~\ref{fig:uniform:genome}\emph{a} shows  the dynamic of the predator mortality distribution for various  initial conditions (marked by different colors). 
In all simulations, sharks have a well-defined phenotype at each time step, so no speciation is observed. 
All distributions, even if with different timescales, tend at long times to converge toward a unique final distribution with a mean close to the average value chosen for the entropic force (dashed line in the figure).
The time required to reach this value diverges quickly (following a power-law with exponent $\sim 2.57$), the farther we prepared the system from the final state.
Note that the exponent of the
power-law slightly depends on the genome size $N$, 
as shown in \figurename~\ref{fig:uniform:genome}\emph{c}.
This kind of dynamic reminds of a system close to a glassy phase~\cite{Kob2000} where the distance between the initial and final phenotypes plays the role of the difference between the quenched and equilibrium temperatures in a typical quenching experiment.
Such behavior provides great insight into the evolutionary dynamics.
Firstly, we observe the role natural selection plays in freezing the system into a meta-stable state for long times. In fact, phenotypes subject to the sole effect of entropic forces would converge to the steady-state exponentially fast (see SI).
Furthermore, we see how the ecosystem and predators, in particular, react to negative mutations. In fact, shark fitness decreases due to the entropic force of the mutations (see SI for the anti-correlation between fitness and shark death-rate). 
If a population has a phenotype distribution whose mean is far from the mean entropy value ($\braket{x}$ in \eqref{eq:q}),  a mutation of the genome will tend to produce an individual with lower fitness than the rest of the population. 
The lowest the fitness the faster this individual will be suppressed by natural selection.  
 

If the fitness worsens only slightly (the mutation is very soft), the specimen will reproduce and that deleterious mutation will remain inside the genome pool. On long timescales, the accumulations of these very soft negative mutations slowly drive the whole population toward the entropic limit. 
This phenomenon is observed in real ecosystems and is known as mutational meltdown\cite{OHTA_1973,Lynch_1995,Grande_Perez_2005}.
The meltdown we observe originates by the even impact genes have on the phenotype. We discuss how it scales with the system carrying capacity in SI.



\figurename~\ref{fig:uniform:genome} clearly show how evolution is not able to improve the fitness of predators.
In fact, although natural selection is freezing the dynamics (strongly reducing the effect of genetic drift) if we wait for sufficient time, sharks will finally increase their death-rate, therefore reducing their fitness.
This behavior is opposite to what is commonly observed in nature, where the combined effect of mutations and natural selection leads to an overall improvement of the fitness on long times. 

This finding highlights the deleterious role of entropy in the evolution of the species. 
We will show in the next section that the capability to improve the fitness can be recovered even in the presence of a strong entropic force if genome variability is considered. This feature alone will be able both to allow predators to evolve and to provide a mechanism for spontaneous speciation.

\subsection{Heterogeneous genome} \label{sec:2}

To include the effect of genome heterogeneity in the simulation, i.e. the uneven impact that different genes have on the phenotype, we inserted a gene-dependent weight (~\eqref{eq:power:law}) in \eqref{eq:pheno}.

\begin{figure}[] 
\includegraphics[width=\columnwidth]{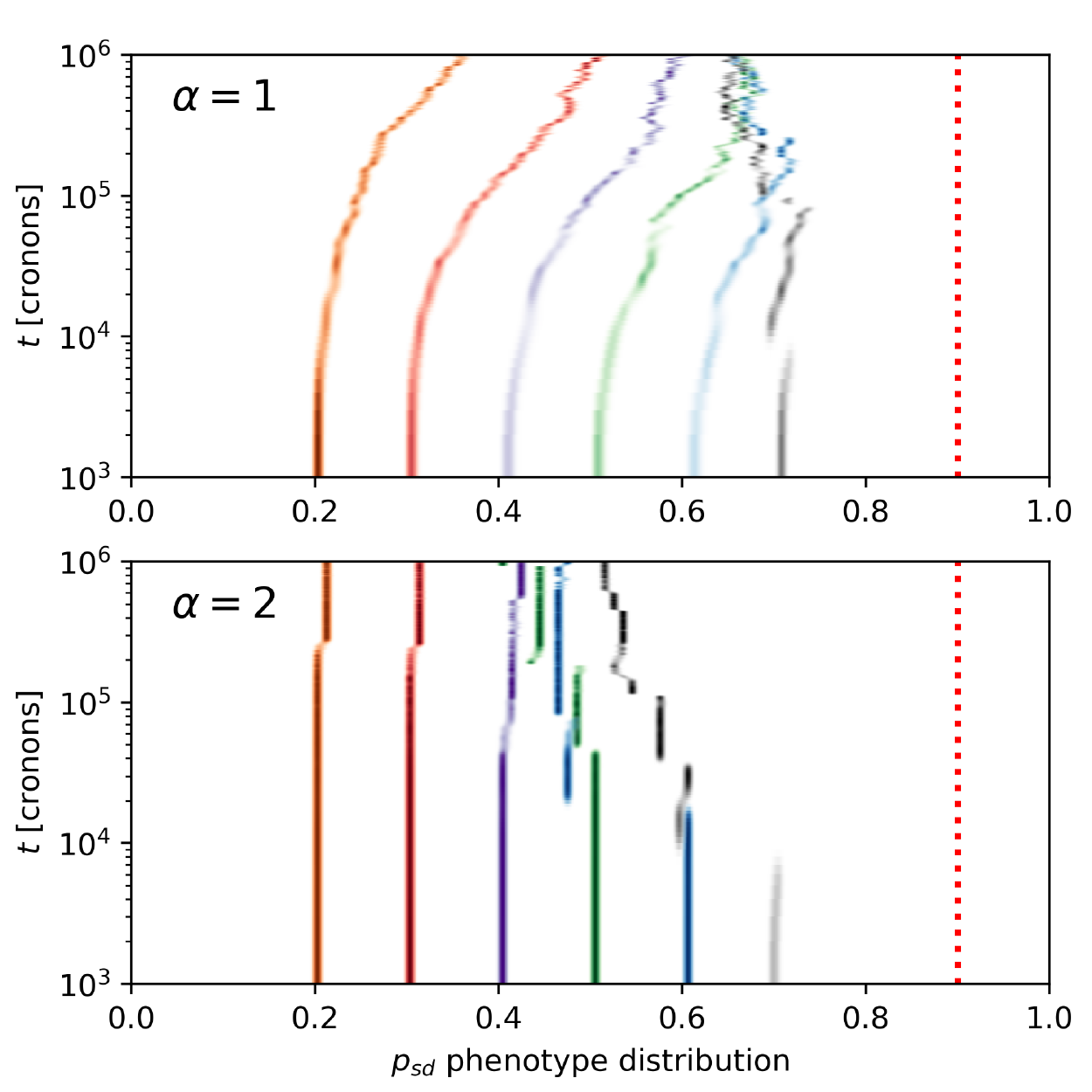}
\caption{\textbf{Evolution of the shark death-probability distribution as a function of time, in presence of a heterogeneous genome.} Heterogeneity allows the specie to evolve to higher average life-time, while this does not occur in presence of a uniform genome (\figurename~\ref{fig:uniform:genome}).  Different colors correspond to different starting values. The two simulation correspond to a different value of the power-law exponent $\alpha$ (see \eqref{eq:power:law}). Both simulations exhibit a qualitative different behaviour if compared to a uniform genome. The less fitted individual can improve their fitness far above the entropic value.}
\label{fig:alpha:genome}
\end{figure}

\figurename~\ref{fig:alpha:genome} depicts the time evolution of the distribution of shark's mortality starting from different initial values  for two different exponent of~\eqref{eq:power:law}, namely $\alpha = 1$ and $\alpha = 2$. 
The evolution is observed with both choices of exponents. Predators improve their expected life-time and the fitness (see SI) much above the equilibrium value obtained in \figurename~\ref{fig:uniform:genome}.  

Positive evolution occurs thank to the combination of two effects; i) the phenotype population is normally quenched (see \secname~\ref{sec:1}) and ii) genes exert a different role on the phenotype manifestation. 
The quenching allows the quasi-species to survive in an out-of-equilibrium situation where common random negative mutations usually kill the individuals.
The species can survive long enough that a very rare, positive, mutation occurs in an important gene (with a big $W_i$), causing a discontinuity in the phenotype distribution and a sudden evolution of the population.
This speciation creates two kinds of predators, with a quantitative different phenotype. The quasi-species which is more fitted to the environment very quickly gets fixed while the less fitted alleles vanish from the gene pool of the population.

Such a rare, discontinuous event is not possible in a uniform genome, where each gene has only a moderate impact on the overall phenotype and a massive number of positive mutations would be required to obtain the same shift ( that is a so rare event that never occurs in practice). Conversely, heterogeneity allows the positive discontinuity in the phenotype to depend on the mutation of a relatively small number of genes, which is much more likely to occur.

\begin{figure*}[t]
\includegraphics[width=14.4cm,height=9.4cm]{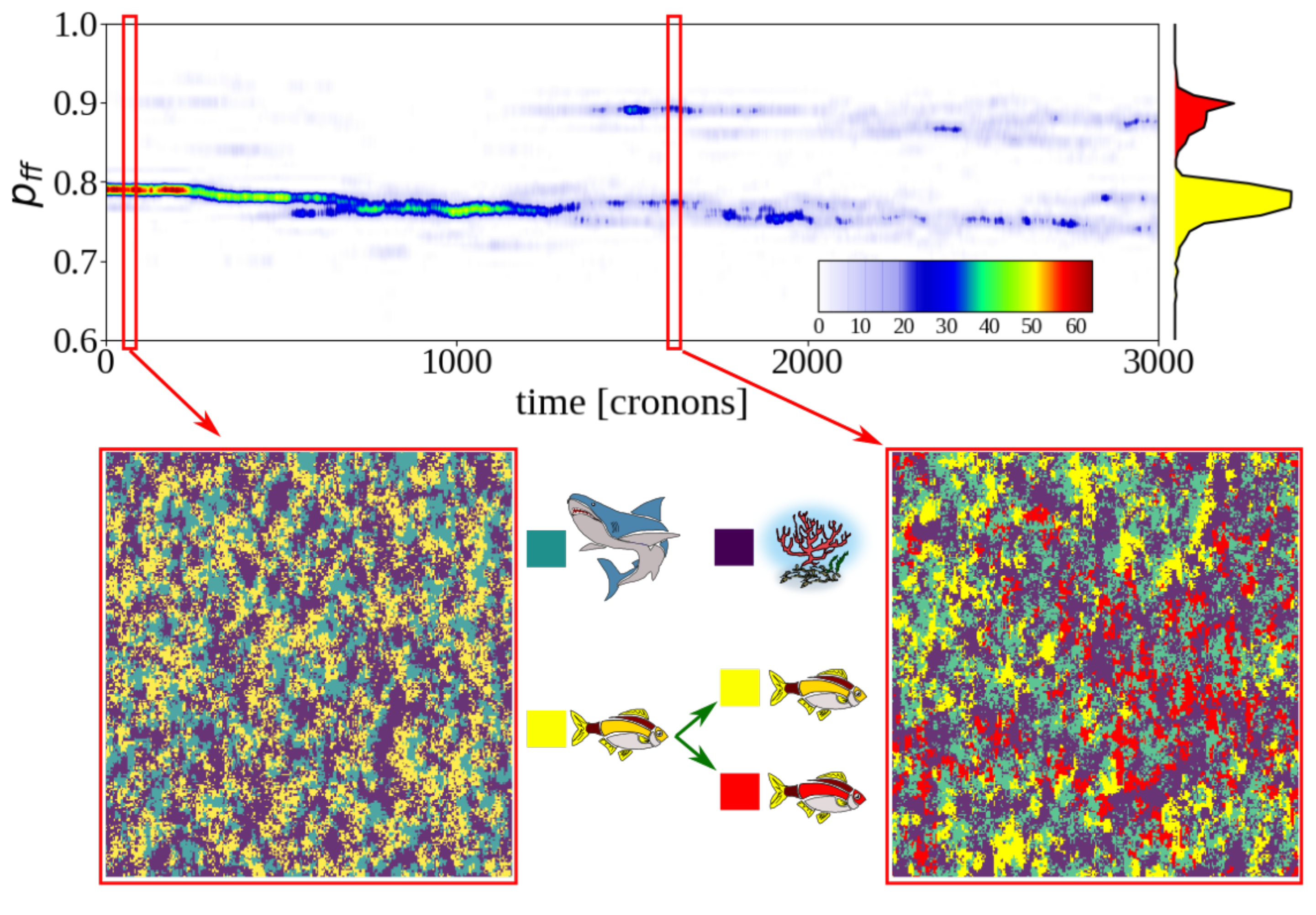}
\caption{\textbf{Outcomes of evolution: speciation.} \textbf{a)} Time evolution of the distribution of fish filiation probability. At time zero, the population of fishes is prepared with a filiation rate of 0.8. After about 1500 chronons, the population splits into two species with a well-separated trait. \textbf{b)} Snapshots of two EvoLat configurations during the evolution. In the left snapshot, only a species of fish is observed. In the right-side one, two fish species have formed. The spatial distribution of sharks differs in the proximity of the two species, spotting that sharks are adopting different hunting strategies.}
\label{fig:speciation}
\end{figure*}

Heterogeneity has another important effect on preys phenotype distribution. In \figurename~\ref{fig:speciation} we show the preys breeding rate as it evolves as a function of time. After a while, the prey population splits into two well-separated traits.
In particular, the two species have a different reproduction rate ($p_f^f$), respectively of $0.78$ and $0.90$, but they share the same fitness in the environment, therefore no one overcomes the other.
This is a very important feature for sympatric speciation, and it is possible only thanks to the presence of the spatial organization of the animals.
In fact, the species with a higher reproduction rate tend to regenerate the shoals faster, being, therefore, more prone to shark predation. This delicate trade-off between higher fertility and the different spatial organization of the shoal that favors predation stabilizes the two species, that can coexist.
In a mean-field scenario, predators would hunt the two species in the same way, and this leads, at a long time, to a supremacy of the prey with the higher reproduction rate.


\section{Discussion}
We simulate the evolution of a minimal prey-predator system. Each species is characterized by a toy genome through which three macro-phenotypes manifest. Providing each specimen with $N$ genes for each phenotype, we showed that the gene relevance (i.e. the weight of each gene according to Eq.~\ref{eq:pheno}) in coding for phenotypes is a key feature for evolution.

To survive, an organism must be robust to deleterious mutations. A convenient choice, the organism can opt for, could be to rely on many genes for the manifestation on one phenotype. If the information for the phenotype is evenly distributed in several genes, then a damage on some gene has the minimum impact on the resulting phenotype.
As a counterpart, we showed that this condition favors the accumulation of detrimental mutations that lead to the mutational meltdown.
Conversely, packing all the phenotype information in few genes enables abrupt variations of phenotype. This prevents the mutational meltdown,
at the cost of reducing the differentiation between individuals and consequently the adaptability of the species. In this case, a drastic change in the environment would provoke a sudden extinction of the species.

Relying on a heterogeneous distribution of information in many genes assures both broad differentiation of a huge genome, and the possibility to have astonishing positive mutations that drive the evolution and prevents the mutational meltdown. 
EvoLat, albeit of representing a minimal model of an ecosystem, reproduces a rich variety of scenarios, upon varying a single parameter.
In fact, tuning the $\alpha$ exponent in Eq.~\ref{eq:power:law}, i.e. assigning different weights to genes associated to the same phenotype, the system exhibits different behaviors.
Indeed, a uniform distribution of weights ($\alpha=0$) leads to a progressive reduction of the fitness due to the accumulation of detrimental mutations (see Fig.~\ref{fig:uniform:genome}). 
On the opposite side, if only one gene encodes for all the traits ($\alpha = \infty$) the mutational meltdown is prevented as species can improve their fitness. However, in this regime, there is no differentiation inside the same population, exposing the species to the threat of sudden environmental changes.
Life lies in between, where the high impact of few genes prevents the mutational meltdown while the bulk of the remaining genes guarantees differentiation.

Notably, fishes provided with a heterogeneous genome respond to the fluctuation of the environment 
and to the natural selection operated by sharks by a sympatric speciation~\cite{FITZPATRICK2008}. 
Spontaneous speciation and in particular sympatric one is rarely reproduced~\cite{Friedman_2013} by models which manage to observe differentiation due to Mendelian inheritance~\cite{AGUIL_E_2011}.
According to our simulations, two key ingredients are important to observe the emergence of sympatric speciations: the heterogeneity in gene relevance and the explicit spatial extension of the simulation. 
Moreover, our minimal model provides a theoretical framework to deal with the everlasting debate about the physical feasibility of evolution.  

The idea that life had evolved by the combined action of mutations in the genome and natural selection of the most fitted individuals is accepted by a vast majority of scientists. Those, who are skeptics, argue that random mutations on the genome should progressively increase the disorder (entropy) and consequently be incompatible with life.   
Indeed, this happens only in case of a uniform genome, where mutations lead the phenotypes toward the entropic values. This argument is, therefore, based on the wrong assumption that all the genes equally contribute to the phenotype.

In our simple model, each part of the genetic sequence influences the phenotype with different weights. 
This is a coarse grain representation of the underlying biological mechanism through which the phenotype is manifested.
In nature, only 2-3\% of DNA is ``coding'' (it can be translated into mRNA and then into proteins). Several studies\cite{biemont2006junk,Prabhakar_2008} have shown that also the non-coding DNA (often referred to as \emph{junk}) can affect the phenotype.
This is a mechanism for strong heterogeneity in how the phenotype manifests from the underlying genetic sequence. 

Overall, in our opinion, the most limiting choices we made were not to include sexual reproduction, correlations in gene expression and modifications in the food chain (a prey cannot become a predator). 
This limits the sources of variation between individuals only to the effect of mutations since both sexual recombination and gene flow are neglected together with the complex correlations between the expression of genes confers to the specimen.
On the other hand, the reduced number of parameters allows one to look for general features of evolution. 

In conclusion, we proved that heterogeneity in gene relevance is a key feature to prevent a mutational meltdown in a species.
Moreover, we showed evidence that spatial correlations are fundamental to account for sympatric speciation.
These findings contribute to disentangling how genomes change to create new species and
provide a step forward in understanding the mechanisms of evolution.


\end{document}


\font\myfont=cmr12 at 40pt
\title{\huge Supplementary Material  \\
 \large Genome heterogeneity  drives the evolution of  species}

\author{Mattia Miotto\footnote{The authors contributed equally to the present work.}}
\affiliation{Department of Physics, Sapienza University of Rome, Rome, Italy}
\affiliation{Center for Life Nanoscience, Istituto Italiano di Tecnologia,  Rome, Italy}

\author{Lorenzo Monacelli\footnotemark[1]}
\affiliation{Department of Physics, Sapienza University of Rome, Rome, Italy}
\affiliation{Center for Life Nanoscience, Istituto Italiano di Tecnologia,  Rome, Italy}

\flushbottom
\maketitle

\thispagestyle{empty}
\newpage

\section*{Entropic forces in farms}\label{app:farm}
Here, we compute the time evolution of a predator population in a ``farm'', i.e. supplying the predator with an infinite source of food, allowing them to reproduce without the effect of natural selection. In this case, each gene of the new generation will have a probability of a mutation given by the mutation rate. Therefore the master equation for the genome of the phenotype $x$ is:
\begin{equation}
    g_{x_i}(t + \delta) = p_\text{mut}\int_0^1 dy \,y q(y) + (1 - p_\text{mut}) g_{x_i}(t)
\end{equation}
\begin{equation}
    \braket{x} = \int_0^1 dy\, y q(y)
\end{equation}
In the limit $\delta\rightarrow 0$ we get:
\begin{equation}
    \frac{d}{dt} g_{xi} = p_{\text{mut}}\left(\braket{x} - g_{x_i}\right)
\end{equation}
\begin{equation}
    g_{x_i}(0) = p_x^0
\end{equation}
\begin{equation}
    g_{x_i}(t) = (p_x^0 - \braket{x}) e^{-p_{mut}t} + \braket{x}
\end{equation}

The overall phenotype of the individual is:
\begin{equation}
    p_x(t) = \sum_{i = 1}^N W_i g_{x_i}(t).
\end{equation}
Since each gene evolves independently in absence of natural selection, we have:
\begin{equation}
    p_x(t) = (p_x^0 - \braket{x}) e^{-p_{mut}t} + \braket{x}
\end{equation}

Therefore, the phenotype converges to the expected value of the entropic forces after a typical time of $1/p_\text{mut}$ independent of the particular initial state $p_x$.

\section*{Mutational Meltdown}
The mutational meltdown is an evolutionary process where deleterious mutations accumulate in time, progressively decreasing the fitness of the population until extinction.

We observe an analogous process in \figurename~2 of the main text.
This process is known to be affected by the carrying capacity. To highlight this effect, we simulated two lattices with different sizes ($L_1 = 256$ and $L_2 = 512$).

We find a (small) increase of relaxation times with the carrying capacity of the system (Fig.~\ref{fig:Lsize}): the higher the carrying capacity, the slower the mutational meltdown. This is in agreement with the theory of mutational meltdown\cite{Grande_Perez_2005,OHTA_1973,Lynch_1995}.

To get an intuitive idea of why the dependence on the carrying capacity is so small, we can do a simple calculation. Natural selection is the force that competes with the accumulation of random deleterious mutations. In a population of size $N$, natural selection acts favoring the most fitted individuals.
A robust population is differentiated, where the individual with the best fitness is distant from the average.
\begin{equation}
    \max \lambda - \braket{\lambda}
    \label{eq:robustness}
\end{equation}

The probability of having the fitness of all individuals below $\lambda_M$ is:
\begin{equation}
P(\lambda < \lambda_M, N) =\left( \int_{-\infty}^{\lambda_M} p(\lambda) \right)^N   
\end{equation}
where $p(\lambda)$ is the distribution of the fitness for each individual. The probability distribution of the maximum fitness among $N$ individuals is
\begin{equation}
\tilde{P}(\lambda_M) = \frac{d}{d \lambda_M} P(\lambda < \lambda_M, N) = N P(\lambda < \lambda_M, N-1) p(\lambda_M)  
\end{equation}

\begin{figure}
\centering
\includegraphics[width=0.6\textwidth]{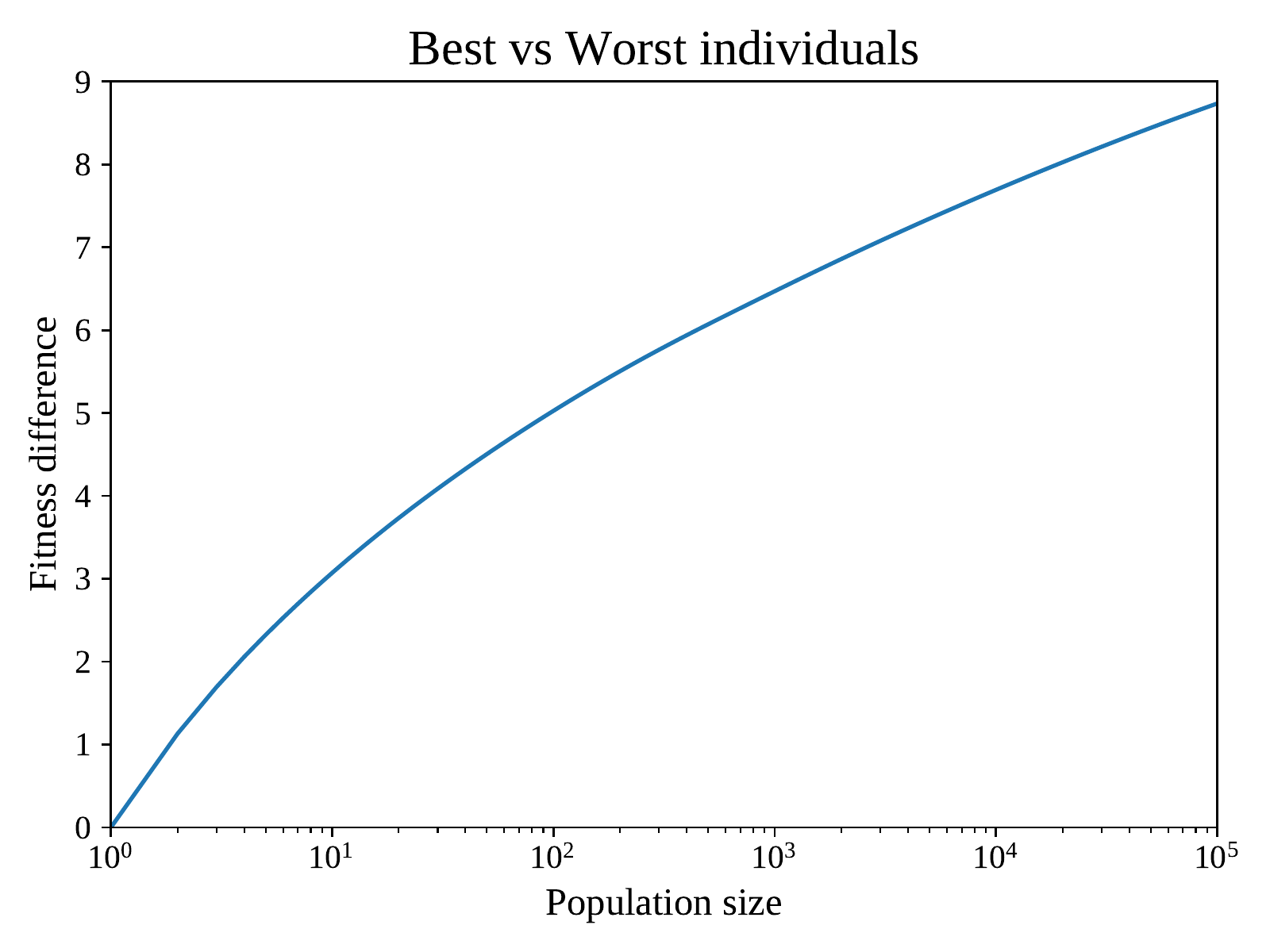}
\caption{Average difference between the most fitted and the mean individual as a function of the population size (robustness to meltdown). The improvement of the robustness of the population increases sub-logarithmic.}
\label{fig:size}
\end{figure}

In figure~\ref{fig:size}, we show the robustness of the population to meltdown \eqref{eq:robustness} versus the size of the population, where $p(\lambda)$ is a Gaussian with unitary variance.
As expected, the robustness of the population increases with its size $N$, however, this grows slower than a logarithm. No surprise that simply quadruplicating the size of the lattice has a small impact on the meltdown.

\begin{figure}
\centering
\includegraphics[width=0.6\textwidth]{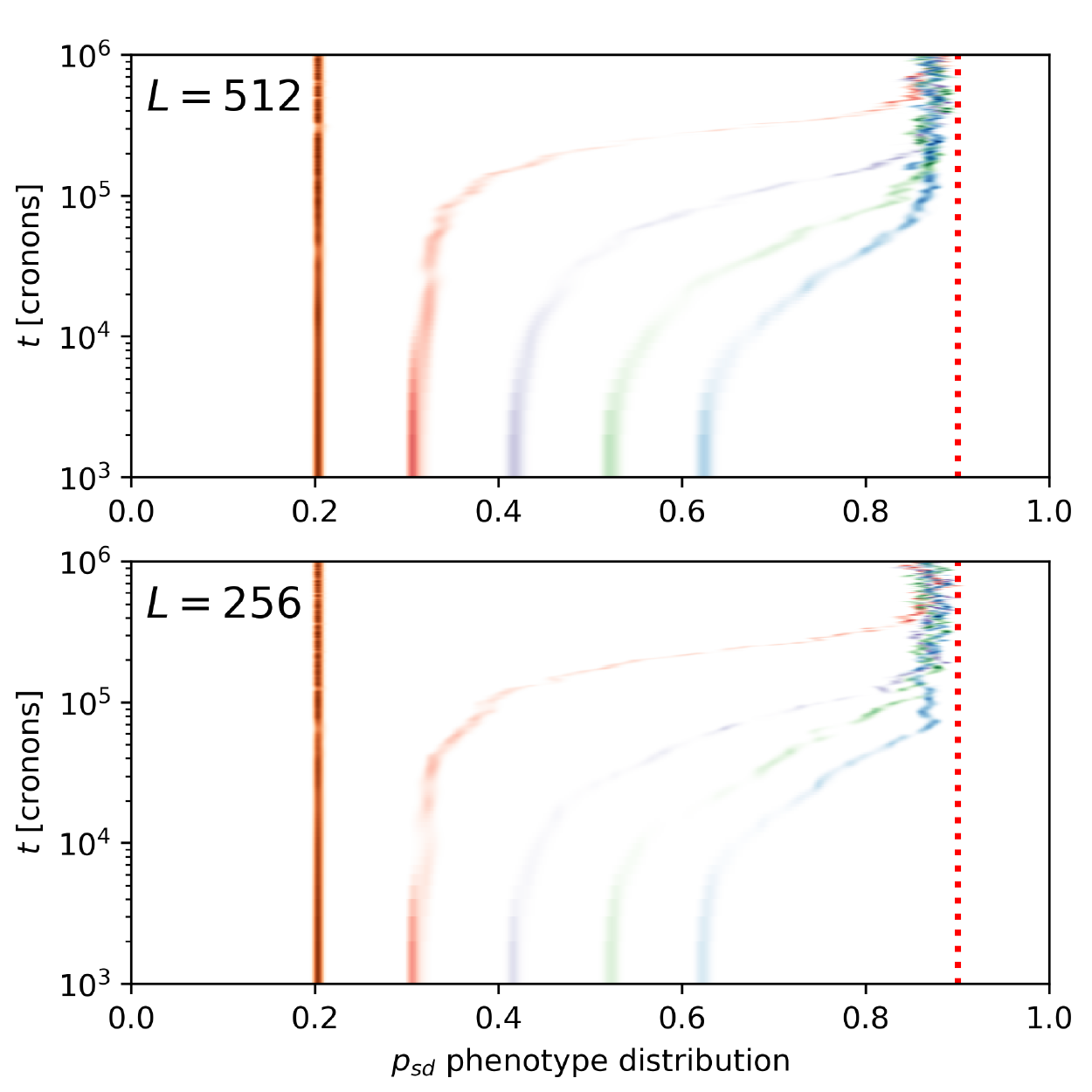}
\caption{\textbf{Time evolution of predators death-rate phenotype after quenching}. The predator relaxation time diverges as a power-law, in a way that resemble the dynamic of glassy systems. Different colors correspond to different simulations performed starting from several values of $p_{sd}$.  The dashed vertical line is the stationary state for predator growing in a breeding farm (no natural selection).\label{fig:uniform:genome} }
\end{figure}

\begin{figure}
\centering
\includegraphics[width=\textwidth]{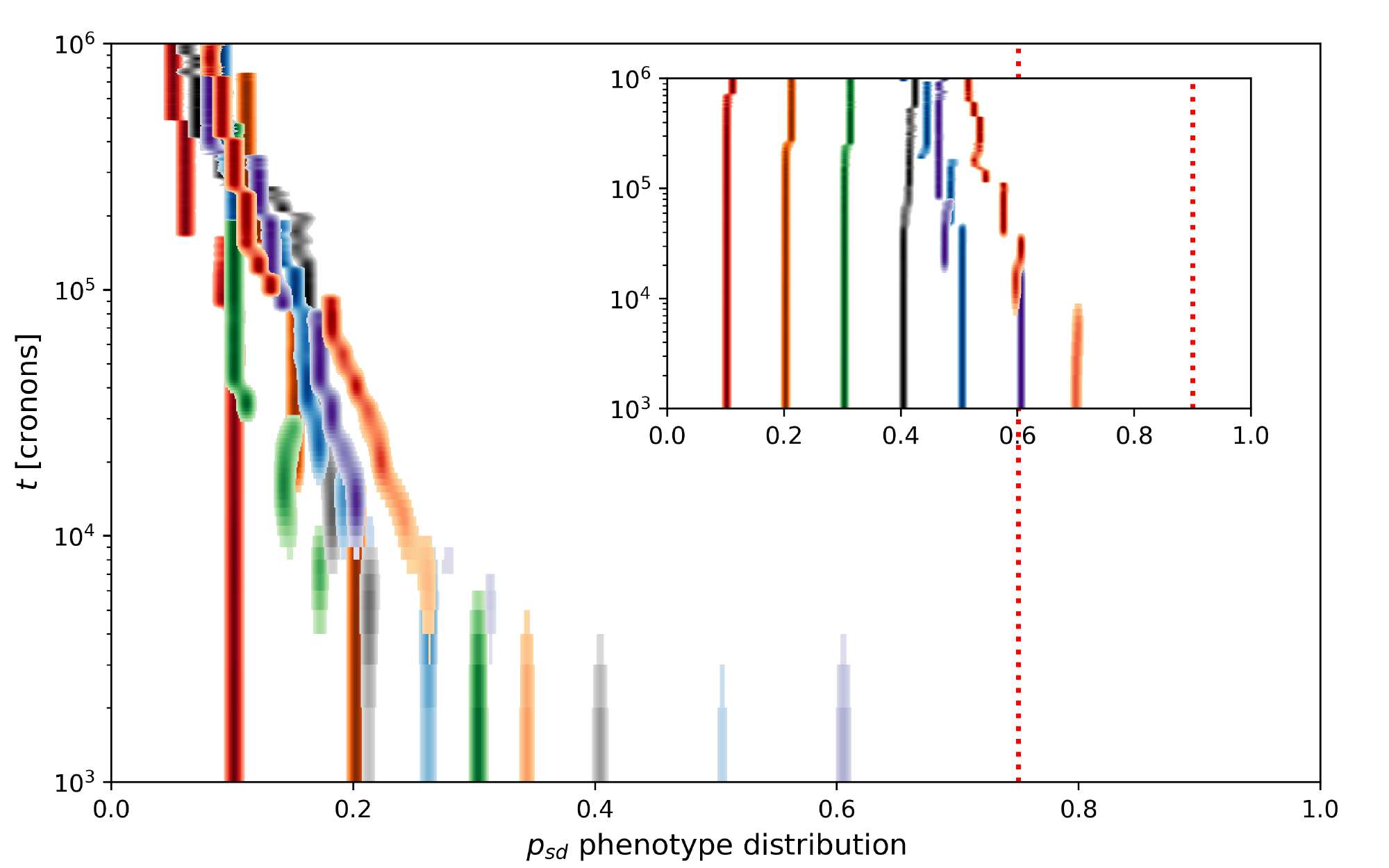}
\caption{\textbf{Evolution of the shark death-probability distribution as a function of time, in presence of a heterogeneous genome.} Heterogeneity allows the specie to evolve to higher average life-time, while this does not occur in presence of a uniform genome (\figurename~\ref{fig:uniform:genome}).  Different colors correspond to different starting values. The less fitted individual can improve their fitness far above the entropic value.\label{fig:Lsize}}
\end{figure}

\begin{figure}
\centering
\includegraphics[width=\textwidth]{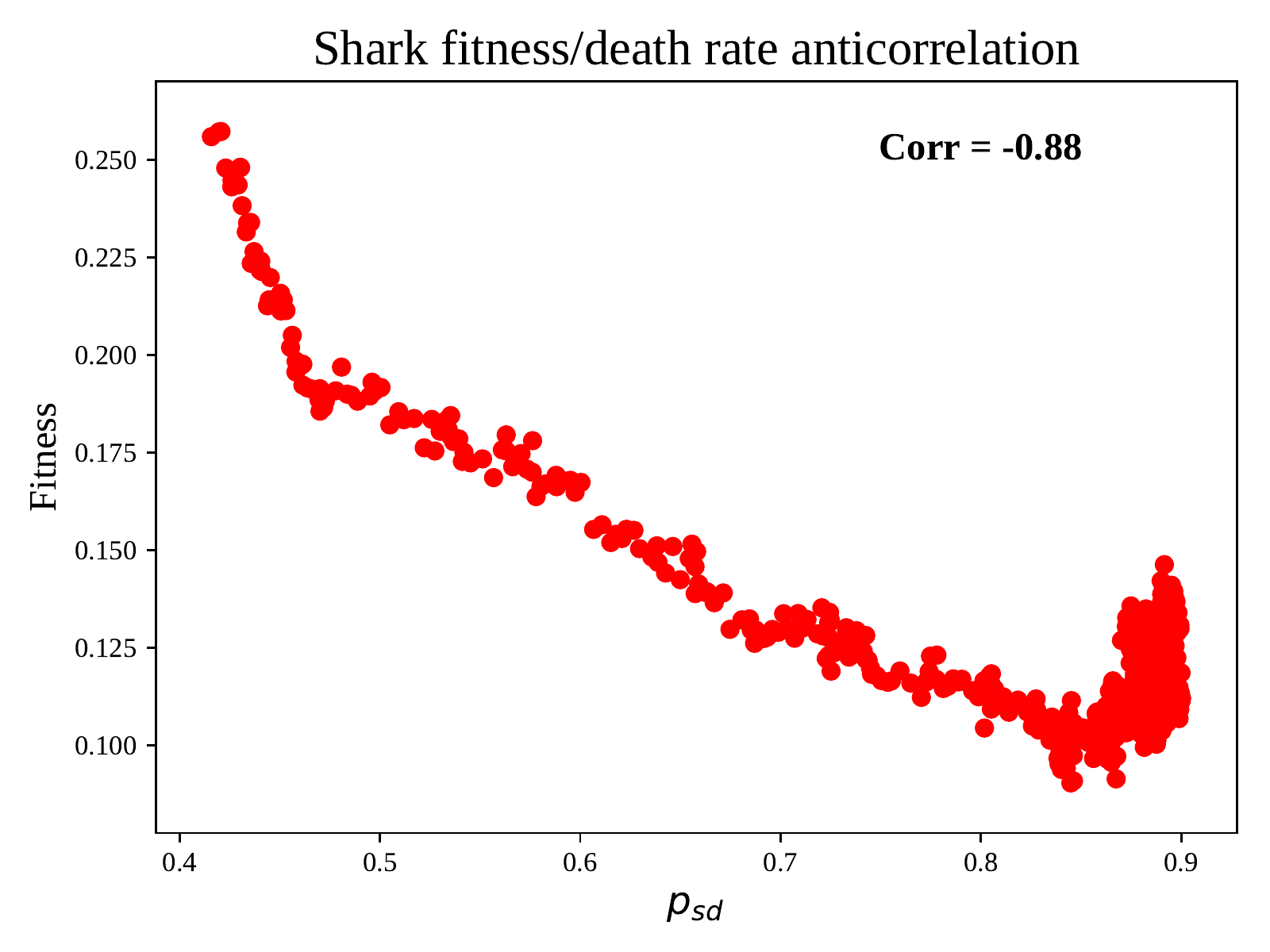}
\caption{\textbf{Anti-correlation between the fitness and the death-rate for predators.} Fitness is defined the average number of individuals in the population.}
\end{figure}
